\documentclass[]{spie}  

\usepackage{amsmath,amsfonts,amssymb}
\usepackage{graphicx}
\usepackage[colorlinks=true, allcolors=blue]{hyperref}

\title{Simulated performance of energy-resolving detectors towards exoplanet imaging with the Habitable Worlds Observatory}

\author[a]{Sarah Steiger}
\author[a]{Laurent Pueyo}
\author[a]{Emiel H. Por}
\author[b]{Pin Chen}
\author[a]{R\'emi Soummer}
\author[a]{Rapha\"el Pourcelot}
\author[c]{Iva Laginja}
\author[b]{Vanessa P. Bailey}
\affil[a]{Space Telescope Science Institute, 3700 San Martin Drive, Baltimore, MD 21218, USA}
\affil[b]{Jet Propulsion Laboratory, California Institute of Technology, 4800 Oak Grove Drive, Pasadena CA 91109, USA}
\affil[c]{LESIA, Observatoire de Paris, Universit\'e PSL, Sorbonne Universit\'e, Universit\'e Paris Cit\'e}

\authorinfo{Further author information: \\
S.S. e-mail: ssteiger@stsci.edu}

\begin{document} 
\maketitle

\begin{abstract}
  One of the primary science goals of the Habitable Worlds Observatory (HWO) as defined by the Astro2020 decadal survey is the imaging of the first Earth-like planet around a Sun-like star. A key technology gap towards reaching this goal are the development of ultra-low-noise photon counting detectors capable of measuring the incredibly low count rates coming from these planets which are at contrasts of $\sim 1 \times 10^{-10}$. Superconducting energy-resolving detectors (ERDs) are a promising technology for this purpose as, despite their technological challenges, needing to be cooled below their superconducting transition temperature ($< 1\mathrm{K}$), they have essentially zero read noise, dark current, or clock-induced charge, and can get the wavelength of each incident photon without the use of additional throughput-reducing filters or gratings that spread light over many pixels. The use of these detectors on HWO will not only impact the science of the mission by decreasing the required exposure times for exo-Earth detection and characterization, but also in a wavefront sensing and control context when used for starlight suppression to generate a dark zone. We show simulated results using both an EMCCD and an ERD to ``dig a dark zone'' demonstrating that ERDs can achieve the same final contrast as an EMCCD in about half of the total time. We also perform a simple case study using an exposure time calculator tool called the Error Budget Software (EBS) to determine the required integration times to detect water for HWO targets of interest using both EMCCDs and ERDs. This shows that once a dark zone is achieved, using an ERD can decrease these exposure times by factors of 1.5--2 depending on the specific host star properties.    
\end{abstract}

\keywords{exoplanets, energy-resolving detectors, detectors, Habitable Worlds Observatory, HWO}

\section{Introduction}
The Astro2020 decadal survey recommends the development of a large IROUV space telescope, now known as the Habitable Worlds Observatory (HWO), with the ambitious goal of imaging $\sim 100$ star systems and talking spectra of $\sim 25$ rocky planets to search for biosignatures that could be indicative of life. This will require the telescope and coronagraph system to disentangle planetary signals that are tens of billions of times fainter than the stars they orbit with planetary count rates on the order of 1-50 photons/hour/$\mu$m around even the nearest Sun-like stars ($\sim$10 pc) \cite{PSG}. Due to these incredibly low count rates, a key technology gap that has been highlighted for HWO are low-noise, photon-counting detectors that will be able to make these measurements with the required signal-to-noise ratio (SNR) in reasonable exposure times. 

Superconducting, energy-resolving detector technologies (ERDs) are of particular interest for this application as, unlike their semiconducting counterparts (e.g. CCDs, EMCCDs, or CMOS detectors), they have no read noise or dark current. They are also ``radiation hard'', where radiation sources such as cosmic rays have no lasting impact on the detector array and can easily be identified and removed from further analysis. Perhaps most importantly, many are also inherently energy-resolving: spectra can be directly obtained without a spectrograph, which typically reduces throughput and spreads out light over many pixels.  

Two of these superconducting ERDs that have risen to the forefront for astronomical applications are microwave kinetic inductance detectors (MKIDs) and transition edge sensors (TESs). For an MKID, each pixel is a superconducting microwave resonator which measures incident  photons through the breaking of Cooper pairs (the charge carriers in the superconductor) which changes the inductance of the material through an effect called the kinetic inductance effect. This causes a shift in the resonant frequency of the pixel where the magnitude of this shift is proportional to the amount of Cooper pairs broken or, equivalently, the energy of the incident photon \cite{2003Day_mkid}. Since each pixel can be designed with its own unique resonant frequency, MKIDs have the benefit of easy multiplexability with large-format IROUV instruments in use for astronomy for over a decade \cite{2013ARCONS, 2018darkness, 2020MEC}. TESs are microcalorimeters where a superconducting film is biased near its superconducting transition temperature allowing for small changes in temperature (caused by a photon event) to create a large resistive signal where the size of this signal is proportional to the photon's energy \cite{2005TES_Irwin}. While TES arrays are not inherently multiplexable, many schemes have been tested to generate large format arrays such as coupling to superconducting quantum interference devices or, more recently, using kinetic inductance current sensors \cite{2002SQUID, 2024paul_kic}. TESs have also achieved better resolving powers in the optical ($R \sim 90$)\cite{TES_r} than MKIDs ($R \sim 50$)\cite{2021deVisser} though fundamentally both MKIDs and TESs can achieve the $R \sim 140$ needed for bio-signature detection with incremental development from the current state-of-the-art. For this work, we will assume that the energy-resolving performance of these detectors has reached these sufficient levels and will be agnostic to either technology since we are focused on their similar high-level noise properties and ability to discern the energy of incident photons with no dispersive optics.

Initial trade studies for HWO have already shown that the use of ERDs can increase exo-Earth yields by up to 30\% over state-of-the-art semiconductor detectors as well as get spectra of $8\times$ as many planets by virtue of constantly collecting spectra \cite{2024Howe}. These studies show that while superconducting detectors add complexity to the mission by needing to be cooled to cryogenic temperatures ($<$ 1 K), their potential benefits could not only outweigh these complexities, but even potentially enable the science goals of the mission.    

\begin{figure}[t!]
    \centering
    \includegraphics[width=\textwidth]{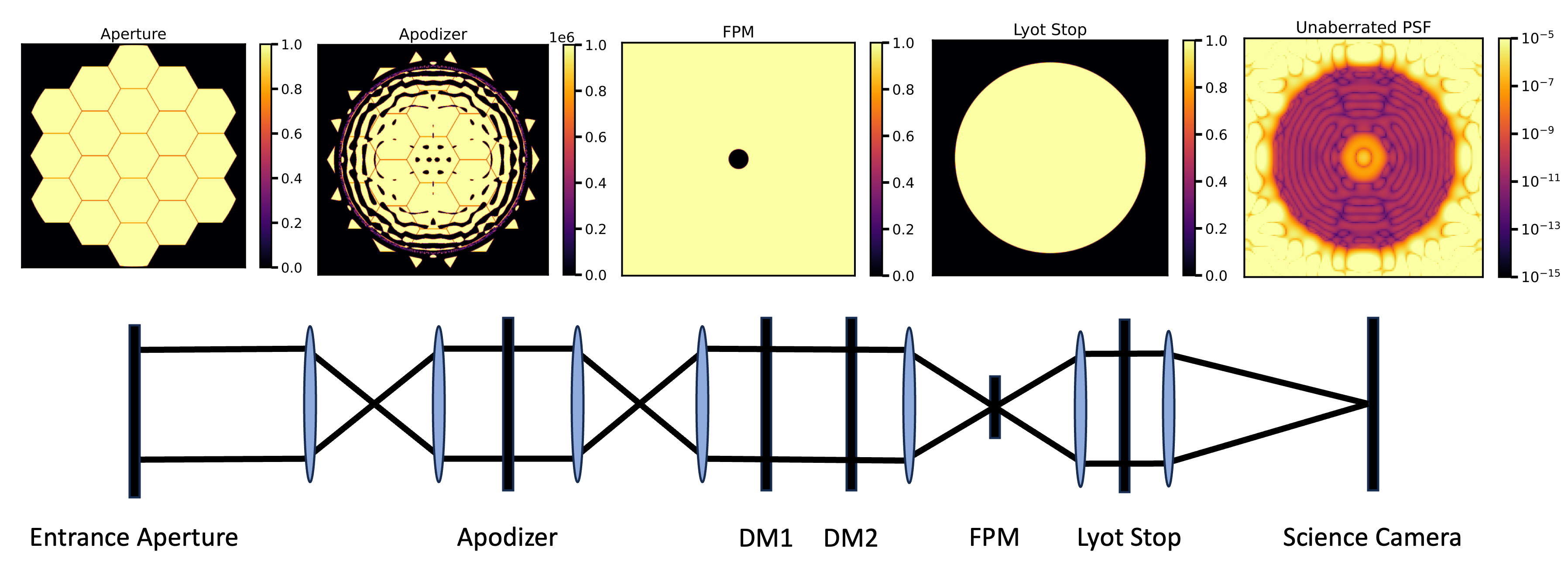}
    \caption{HWO EAC1-like optical model used for all WFS\&C simulations.}
    \label{fig:optical_model}
\end{figure}

In addition to the increase in exo-Earth candidate yield and incidental spectra when using an ERD, in recent years it has become clear that for HWO to achieve its science mission goals, the whole observatory \textit{system} (from the telescope, to the coronagraph, to the detector) will need to be optimized together to maximize science output. In this work we will explore how ERDs can be integrated with a coronagraphic system and improve the performance of ``digging a dark zone'' in a wavefront sensing and control (WFS\&C) loop to reduce overheads and allow more mission time to be dedicated to performing science. 

We start by outlining our optical model and simulation scheme in Section \ref{sec:optical_model}, including the detector implementations and assumptions for both EMCCDs and an ERD. We then present the results of the efficiency of using an ERD or EMCCD to dig a dark zone in Section \ref{sec:results}. We end in Section \ref{sec:etc_study} with a brief case study using an exposure time calculator tool called the Error Budget Software (EBS) to further demonstrate the increased efficiency of using an ERD for science exposures after a dark zone has been achieved.   

\section{Optical Model and Simulation}
\label{sec:optical_model}

\subsection{Telescope Model}

The optical model used for this work was generated using the same infrastructure as the optical model for the High-contrast Imager for Complex Aperture Telescopes (HiCAT) testbed \cite{2018hicat_software, 2022Hicat8, SoummerSPIE2024} at the Space Telescope Science Institute (STScI). To match the notional exploratory analytic case 1 (EAC1) design for HWO, this existing optical model was updated with the parameters found in Table \ref{table:sim_params} and illustrated in Figure \ref{fig:optical_model}.

For this study, an apodized-pupil Lyot coronagraph (APLC) design was chosen for starlight suppression. Specifically, we used the 2-Hex design from Nickson et. al (2022) \cite{2022_APLC_optimization} with 0.2\% tolerance to Lyot Stop misalignment, see their Figure 9. We also adopted the same focal-plane mask (FPM) and Lyot stop specifications as is found in that paper. In the presence of no additional aberrations, this design has a mean contrast of $3.6 \times 10^{-11}$ from $3.5-12 \lambda/D$.

In addition to the optical model, we also made use of HiCAT's existing simulator to perform our WFS\&C experiments. This simulator uses the open-source Control and Automation for Testbeds Kit 2 (catkit2) \cite{por_2024_catkit} as its backend, which allows for high-fidelity simulations of all of the mechanical components of a high-contrast imaging system. Relevant to this work, this includes realistic camera modules, and deformable mirror (DM) control channels. 

\begin{figure}
    \centering
    \includegraphics[width=0.5\textwidth]{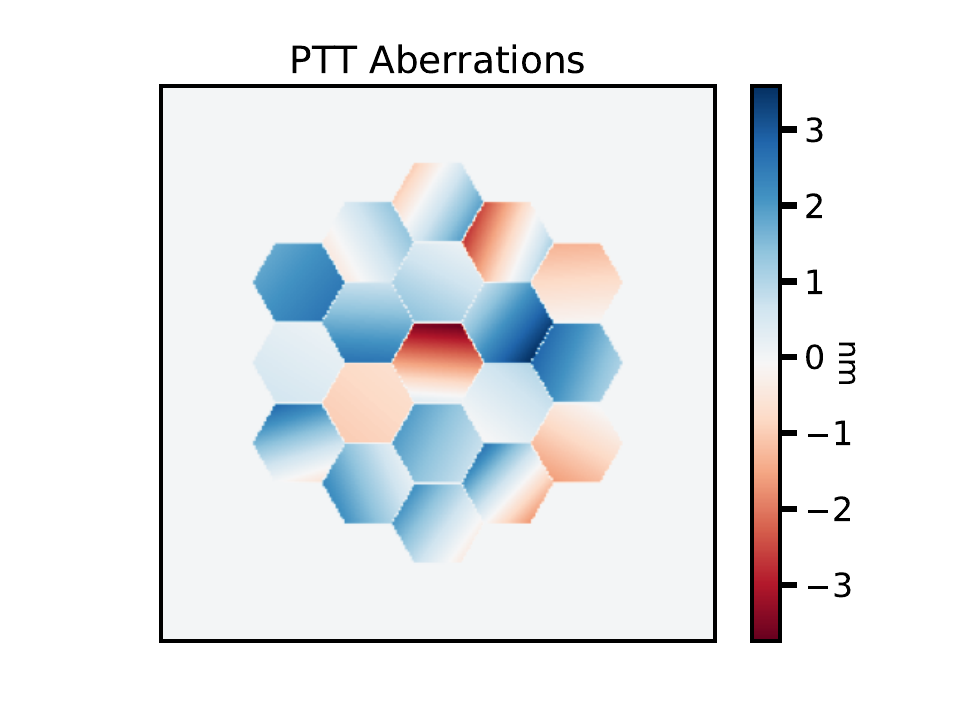}
    \caption{Input piston, and tip/tilt (PTT) aberrations (surface, in nm) to the primary mirror in the optical model to achieve a starting contrast of $\sim 1 \times 10^{-6}$.}
    \label{fig:PTT}
\end{figure}

In order to simulate a more realistic observing scenario, we introduced piston and tip/tilt (PTT) aberrations onto the segmented primary mirror (see Figure \ref{fig:PTT}) to achieve a starting contrast of  $1.2 \times 10^{-6}$ which was the starting place from which all WFS\&C experiments were run.

\begin{table}[]
\caption{Optical model parameters.}
\centering
\begin{tabular}{||l l l||} 
 \hline
 Parameter & Value & Description\\ [0.5ex] 
 \hline\hline
 &&\textbf{General Parameters} \\
 $\lambda_c$ & 640 nm & Central wavelength of the bandpass \\ 
 
 $\Delta\lambda$ & 60 nm & Bandwidth \\ 

 $D$ & 7.26 m & Circumscribed diameter of telescope \\ 

 $D_{ins}$ & 6.0 m & Inscribed diameter of telescope \\

 $N_{s}$ & 19 & Number of segments  \\

 s & 1.67 m & Segment size (point-to-point)  \\

 $d_{LS}$ & 0.98 $D_{ins}$ & Lyot stop diameter  \\

 $d_{FPM}$ & 3.5 $\lambda/D$ & Focal plane mask size  \\

 $N_{act}$ & 48 x 48 & Number of actuators per DM \\ 
 \hline
 && \textbf{HWO EMCCD \cite{2019LUVOIR, 2020HabEx}}\\
 
 CIC & $1.3 \times 10^{-3}  \; \mathrm{e}^{-} \mathrm{pix}^{-1} \mathrm{frame}^{-1}$ & Clock-induced charge \\
 RN & $0  \; \mathrm{e}^{-} \mathrm{pix}^{-1} \mathrm{read}^{-1}$ & Read noise\\
 $\zeta$ & $3.0 \times 10^{-5}  \; \mathrm{e}^{-} \mathrm{pix}^{-1} \mathrm{s}^{-1}$ & Dark current\\ 
 QE & 0.9 & Quantum efficiency \\
 \hline
 && \textbf{Roman EMCCD (BOL) \cite{roman_detect_params}}\\
 CIC & $1.6 \times 10^{-2}  \; \mathrm{e}^{-} \mathrm{pix}^{-1} \mathrm{frame}^{-1}$ & Clock-induced charge \\
 RN & $1.1 \times 10^{-4}  \; \mathrm{e}^{-} \mathrm{pix}^{-1} \mathrm{read}^{-1}$ & Read noise\\
 $\zeta$ & $1.4 \times 10^{-5}  \; \mathrm{e}^{-} \mathrm{pix}^{-1} \mathrm{s}^{-1}$ & Dark current\\ 
 QE & 0.75 & Quantum efficiency \\ 
\hline
  && \textbf{ERD}\\
 CIC & 0 $\mathrm{e}^{-} \mathrm{pix}^{-1} \mathrm{frame}^{-1}$ & Clock-induced charge \\
 RN & 0 $\mathrm{e}^{-} \mathrm{pix}^{-1} \mathrm{read}^{-1}$ & Read noise\\
 $\zeta$ & 0 $\mathrm{e}^{-} \mathrm{pix}^{-1} \mathrm{s}^{-1}$ & Dark current\\ 
 QE & 1.0 & Quantum efficiency \\ [1ex] 
 \hline
\end{tabular}
 \label{table:sim_params}
\end{table}

\subsection{WFS\&C Strategy}

In order to image faint Earth-like planets at contrasts of $10^{-10}$, diffracted starlight in the focal plane caused by non-common path errors and telescope imperfections needs to be corrected for and removed in the region of interest. This process is typically called ``digging a dark zone'' and uses DMs to sense and remove unwanted stellar contamination at the location of potential planets.  

To dig our dark zone, we used electric field conjugation (EFC)\cite{2007EFC} with pairwise probing. In this experiment we used four sets of single actuator probes with wavefront error amplitudes of 10~nm, where each probe set contains a positive and negative probe for a total of 8 probe images per EFC iteration -- for more details, see Soummer et. al. (2024)\cite{SoummerSPIE2024}.


\subsection{Stellar Flux Calibration}

Count rates were used to mimic the observing of a star with a V-band magnitude of 8 and were calibrated using the exoscene\footnote{https://github.com/nasa/exoscene} library, assuming an optimistic total system throughput of 50\%. These count rates were also cross-checked using the EXOSIMS \cite{2017EXOSIMS} package and found to agree to within a factor of 2. A magnitude 8 star was chosen to reflect the dimmer stars found in the NASA Exoplanet Exploration Program (ExEP) mission star list for the Habitable Worlds Observatory \cite{2024ExEp_catalog} and a typical V-band magnitudes as found in the the Habitable Worlds Observatory preliminary input catalog (HPIC) \cite{2024HPIC}.

\subsection{Detector Implementations}
\label{sec:detector_model}

We made use of the EMCCDDetect\footnote{https://github.com/roman-corgi/emccd\_detect} package for simulating EMCCD images given a ``perfect'' input wavefront which is based on work by Nemati 2020b \cite{2020Nematib}. This package adequately handles the main sources of noise in the EMCCD including clock-induced charge (CIC), dark current ($\zeta$), and read noise (RN). 

We assume three different detector scenarios whose properties are summarized in Table \ref{table:sim_params}. These reflect the nominal HWO detector as recommended by the LUVOIR and HabEx reports (HWO EMCCD), a detector that has the same properties as the detector in the coronagraphic instrument on the Nancy Grace Roman Space Telescope at its beginning-of-life performance (Roman EMCCD BOL), and a noiseless ERD (ERD). 

In this study we additionally assume an optimistic quantum efficiency (QE) of 1.0 for the ERD and 0.9 for the HWO EMCCD. We chose to ignore the effects of cosmic rays though think this is an interesting path for future work due to the fundamentally different ways that cosmic rays appear in and ERD compared to an EMCCD. Finally, Poisson noise is added to each image using the HCIPy \cite{por2018hcipy} \texttt{large\_poisson} utility.

\subsubsection{Modeling of an Integral Field Unit (IFU)}
Following the procedure outlined in Nemati et. al. 2020a\cite{2020nemati_coronagraphs}, to model an IFU a sampling of 4 pixels per lenslet is chosen. When calculating a broadband WFS\&C solution, only a small number of spectral elements ($N_{spec}$) over the bandpass is desired to maximize throughput and reduce necessary computation time. In general, 
\begin{equation}
    N_{spec} = R \cdot BW,
\end{equation}
where $R$ is the spectral resolution of the IFU and $BW$ is the instrument bandwidth. It then follows that the number of pixels light would be spread over ($N_p$) is 

\begin{equation}
    N_p = N_{spec} \cdot l_{samp},
\end{equation}
where $l_{samp} = 4$ is the lenslet sampling. 
If a dedicated IFU with an $R = 30$ and $BW = 10\%$ is flown on HWO to be used primarily for WFS\&C, then $N_{spec} = 3$ and the factor of 4 to account for the IFU lenslet sampling would be the only multiplicative factor on the detector noise that needs to be considered. This resolution however is too small to detect key bio-signatures and so a second higher resolution ``science-grade'' IFU would be needed. 

If instead a science grade IFU with an $R = 140$ is used for WFS\&C then the detector noise contribution will be much higher. In this case $N_{spec} = 140 \cdot 0.1 = 14$ with the light being spread out over $N_{pix} = 14 \cdot 4 = 56$ pixels. Rebinning this to the desired spectral resolution ($R = 30$) to maximize throughput would mean that the pixel noise contribution per spectral element is $\frac{140}{30} \cdot 4 \sim 19$ pixels. In this work, we consider this science-grade IFU ($R = 140$) case where the detector noise is multiplied by a factor of 19. It is important to note that the ERD does not have these considerations since it can effectively be thought of as an IFU with a variable $R$ and does all wavelength binning in post-processing.

\section{Results}
\label{sec:results}

In this work, we explored four WFS\&C strategies and their impacts on the overhead to dig a dark zone and begin taking science observations of exo-Earth candidates. They are (1) using an imaging EMCCD (i.e., no dispersive optics to get wavelength information), (2a) Using an HWO-like EMCCD as part of a science-grade IFU, (2b) using a Roman-like EMCCD as part of a science-grade IFU, and (3) using an ERD.   

\begin{figure}
    \centering
    \includegraphics[width=0.6\textwidth]{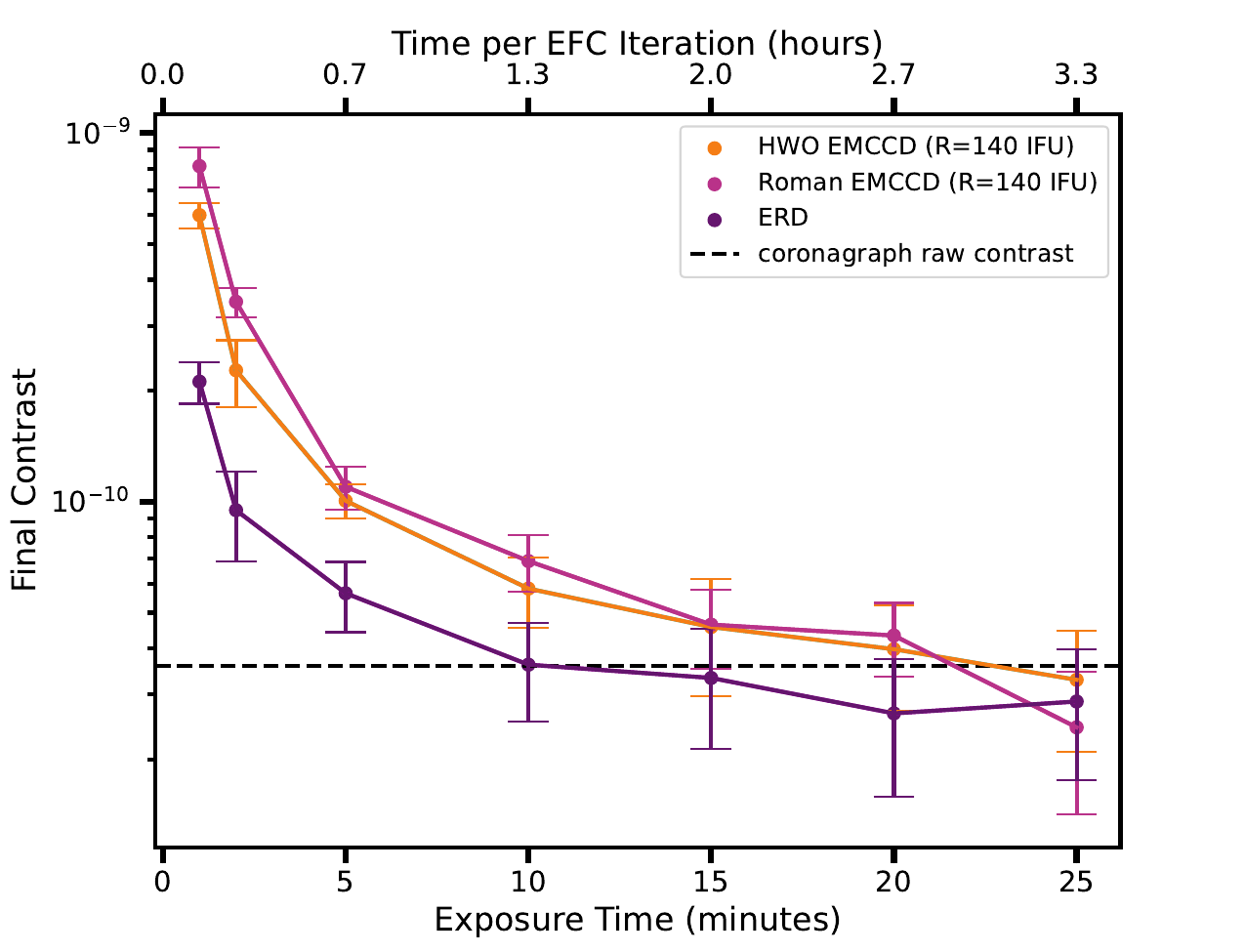}
    \caption{Final contrast achieved as a function of exposure time per probe image (bottom axis) and total time per EFC iteration (top axis) for an HWO EMCCD (orange), a Roman EMCCD  (pink), and an ERD (purple).}
    \label{fig:hwo_erd_contrast}
\end{figure}

In case (1) there is an obvious overhead increase due to the need to take exposures at multiple wavelengths to calculate a broadband control solution. For example, consider needing $t_{exp} = 1000$ seconds of exposure time per spectral bin to acquire enough signal on the detector (at contrasts of $10^{-10}$) with $N_{\lambda} = 3$ spectral bins over the bandpass. If pairwise probing is used to sense the electric field then it is reasonable to apply between 2-4 sets of probes (pessimistically let us set $N_p = 8$ probes total) per wavelength bin. This results in a total exposure time $T = t_{exp} \cdot N_{\lambda} \cdot N_{p} = 1000\cdot3\cdot8 = 24000$ seconds (6.6 hours) spent sensing the electric field for only a single EFC iteration. 

Using an ERD or IFU, one broadband image per probe would be used instead which, assuming a flat spectrum over the bandpass, results in needing the same exposure time per probe ($t_{exp} = 1000$ s) but with the added benefit of getting all wavelengths simultaneously. This means that only one set of probe images is needed and so $T = t_{exp}\cdot N_{p} = 1000\cdot8 = 8000$ seconds (2.2 hours) per iteration which is a factor of $N_{\lambda}$ less than the imaging case (see also Table \ref{table:time_to_contrast}). It is worth mentioning that for this application an ERD would only require a modest energy resolution of $R \sim \frac{E}{\Delta E} = \frac{600 \mathrm{nm}}{20 \mathrm{nm}} = 30$ to resolve 3 spectral bins in 10\% bandpass at 600~nm for which both MKIDs and TESs have already surpassed.\cite{2021deVisser, TES_r}


\begin{figure}[t!]
    \centering
    \includegraphics[width=0.57\textwidth]{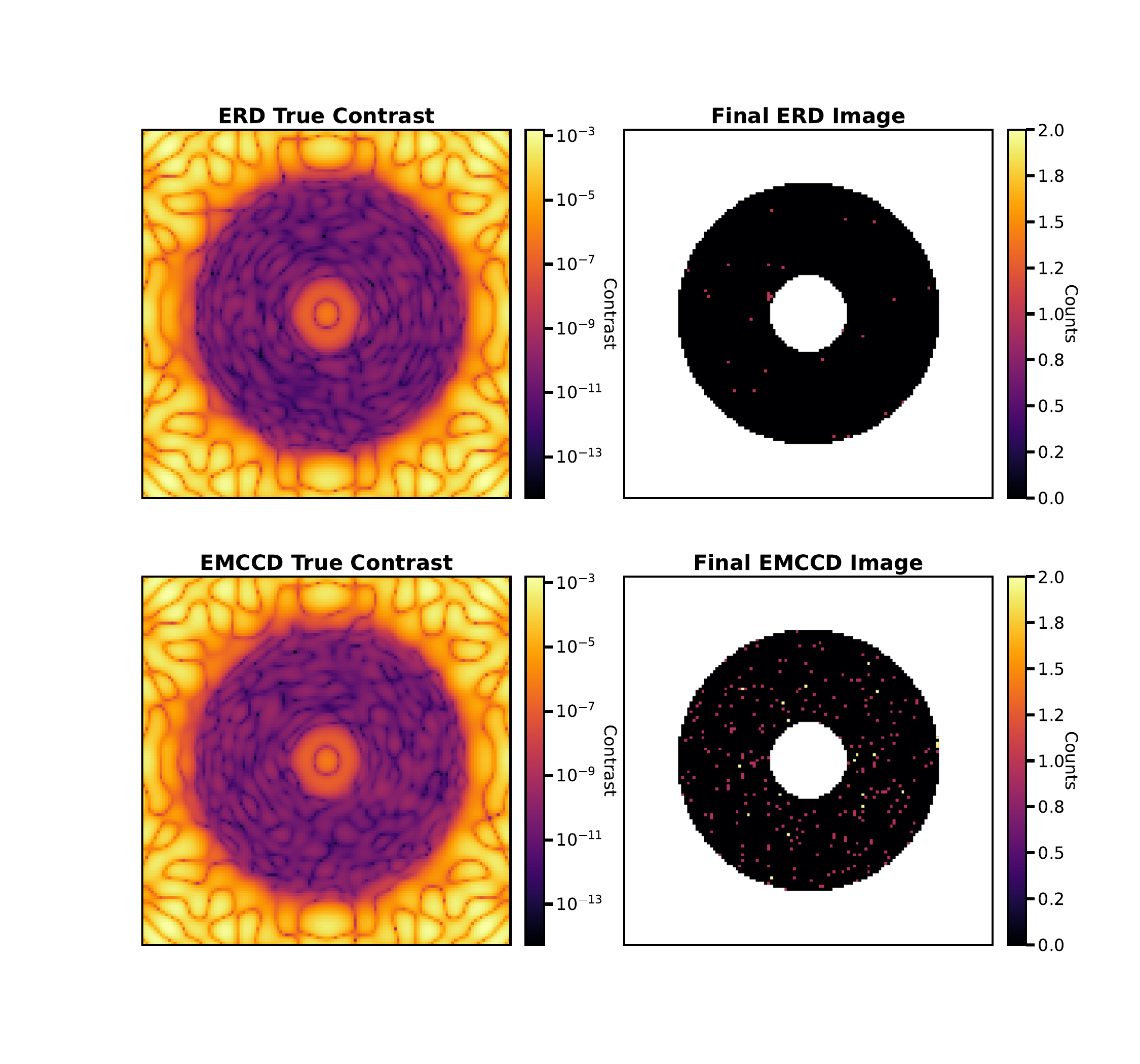}
    \caption{Left: Contrast images applying the final DM solution to the ideal (noiseless) optical model for an ERD (top) and EMCCD (bottom). Right: Final camera images at that same base contrast showing the number of counts present in the dark zone of the final images for the ERD (top) and EMCCD (bottom). These images are for $t_{exp}$ = 5 minutes.}
    \label{fig:final_images}
\end{figure}

Cases (2a), (2b), and (3) are all sensitive to wavelength information and so are instead differentiated by how the detector noise impacts the required exposure times needed to properly sense the electric field or, equivalently, how well the electric field can be sensed and corrected for a given fixed exposure time. In order to test these cases,  we ran WFS\&C simulations using the detector models described in Section \ref{sec:detector_model} and compared the final dark zone contrast as a function of exposure time per image ($t_{exp}$) for an HWO-like EMCCD, Roman-like EMCCD, and ERD. The results are found in Figure \ref{fig:hwo_erd_contrast}. Here the final contrast was determined by taking the final DM solution for each scenario, applying it to the optical model, and then calculating the mean contrast in the dark zone using an entirely noiseless detector model. A noiseless detector model is used here since the final images using the noisy detector models contain between 0--2 counts per pixel for all cases. Therefore, using the normal method for determining contrast for these noisy models (taking the mean of all the values in the dark zone) would not be appropriate. At these count rates, using this method would result in the pixels containing no photons artificially deflating the measured contrast -- see Figure \ref{fig:final_images}.

\begin{figure}[t!]
    \centering
    \includegraphics[width=0.7\textwidth]{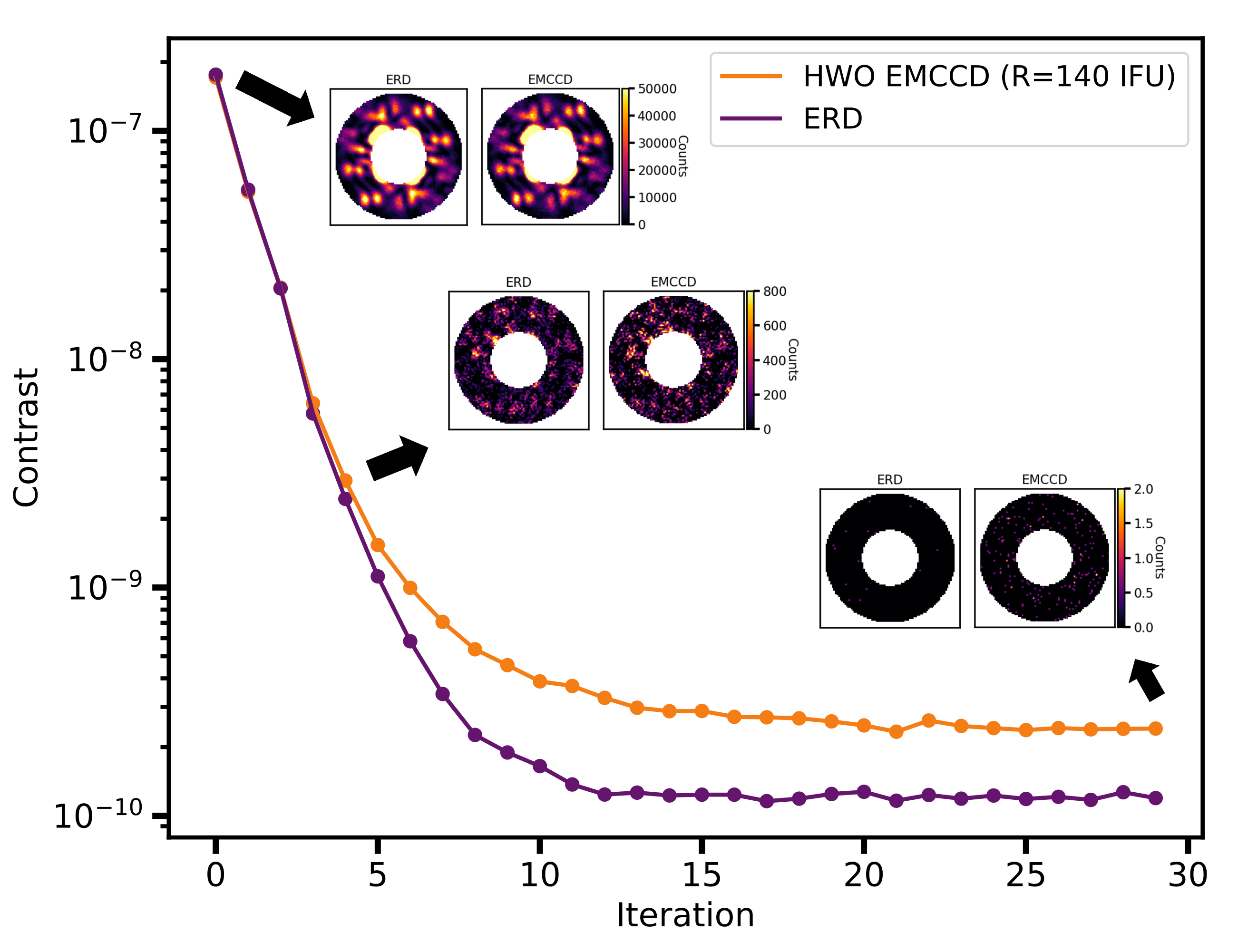}
    \caption{Contrast vs. iteration for the HWO EMCCD (orange) and ERD (purple) with the camera images at select iterations inset. This data is for $t_{exp}$ = 2 minutes.}
    \label{fig:contarst_vs_iter}
\end{figure}

In Figure \ref{fig:hwo_erd_contrast}, it can be seen that at short exposure times, the ERD outperforms both EMCCD scenarios by about a factor of 2 until the point at which the exposure time is long enough that the detector noise threshold is surpassed and the two cases converge. Figure \ref{fig:contarst_vs_iter} shows the contrast vs. EFC iteration for the $t_{exp}$ = 2 minutes case with insets showing the coronagraphic camera images for select iterations. At the fifth iteration where the two curves start to diverge, it can be seen that the additional noise in the EMCCD image is beginning to be on par with the speckle signal likely leading to less efficient sensing of the electric field at these count rates. For a summary of how these scenarios effect the total overall time to achieve a given dark zone contrast, see Table \ref{table:time_to_contrast}.

\begin{table}[]
\vspace{0.5cm}
\caption{Time to DZ Contrast}
\centering
\begin{tabular}{|| l | l l l l||} 
 \hline
 Contrast & 5e-10 & 3e-10 & 1e-10 & 8e-11 \\ 
 \hline\hline
 ERD & $1_{(1.3 \mathrm{h})}$ & $1_{(1.9 \mathrm{h})}$ & $1_{(8.7 \mathrm{h})}$ & $1_{(9.3 \mathrm{h})}$\\ 
 HWO EMCCD (R=140 IFU) & $2.1 \times _{(2.7 \mathrm{h})}$ & $1.9 \times _{(3.7 \mathrm{h})}$ & $1.3 \times _{(11.3 \mathrm{h})}$ & $2.0 \times _{(18.7 \mathrm{h})}$\\ 
 Roman EMCCD (R=140 IFU) & $2.5 \times _{(3.2 \mathrm{h})}$ & $3.2 \times _{(6 \mathrm{h})}$ & $1.8 \times _{(16 \mathrm{h})}$ & $2.0 \times _{(18.7 \mathrm{h})}$\\ 
 Imaging EMCCD & $6.8 \times _{(8.8 \mathrm{h})}$ & $8.8 \times _{(16.8 \mathrm{h})}$ & $4.6 \times _{(40 \mathrm{h})}$ & $9.0 \times _{(84 \mathrm{h})}$\\  [1ex] 
 \hline
\end{tabular}
\vspace{0.2cm}
 \caption{Time needed to reach a given dark zone (DZ) contrast for different detector WFS\&C scenarios. The numbers are all normalized to the ERD with the multiplicative factors denoting how much longer each scenario would take as compared to an ERD. The ``true'' times to reach the given contrast are found in the subscripts though it should be noted that these ``true'' values will be highly sensitive to particular choices in simulated observing parameters such as stellar magnitude or instrument throughput.}
 \label{table:time_to_contrast}
\end{table}

\section{ETC Case Study: Detector Noise}
\label{sec:etc_study}

Clock induced charge (CIC), dark current ($\zeta$), and read noise (RN) are the three main sources of detector noise that are modeled by yield codes and exposure time calculators (ETCs) to determine the time-to-SNR for an observation, i.e. how long one needs to observe a target to reach the desired SNR for detection or characterization. As part of the coronagraph technology roadmap (CTR) group funded by the ExEP Office, an ETC was developed to study the effects of stability, wavefront error, and wavefront sensing and control systems on the expected exposure times for exo-Earth candidate detection and characterization called the Error Budget Software (EBS)\footnote{https://github.com/chen-pin/ebs}. EBS is fundamentally a wrapper for another ETC and yield code called EXOSIMS\cite{2017EXOSIMS} which it uses to perform all of its backend calculations.

\begin{figure}[h!]
    \centering
    \includegraphics[width=0.8\textwidth]{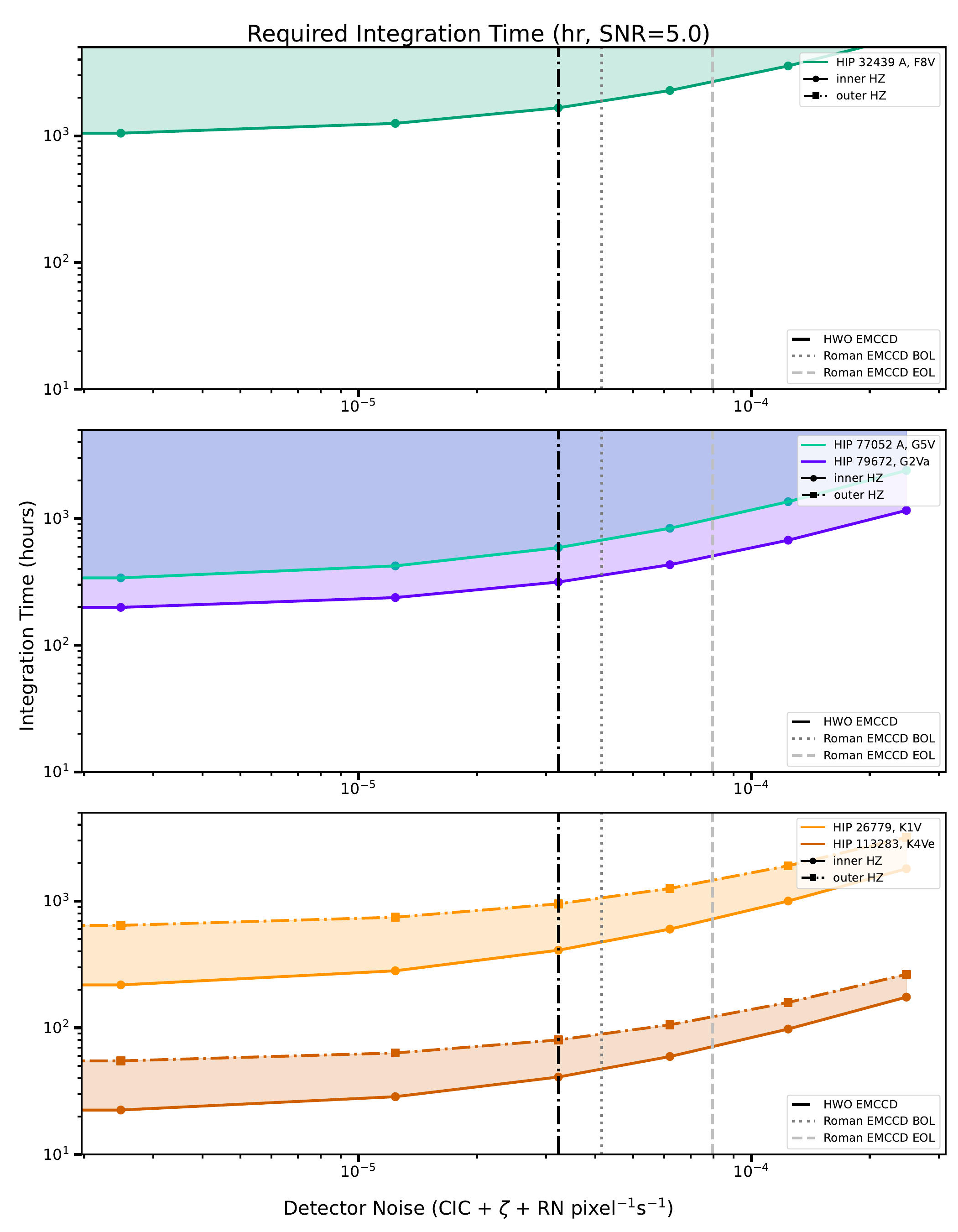}
    \caption{Integration times required to reach a signal-to-noise ratio of 5 for H$_2$O detection ($\lambda$=1000 nm, R=140) for 5 fiducial HWO target stars, one F spectral type (HIP 32439, top), two G spectral types (HIP 77052 and HIP 79672, middle), and two K spectral types (HIP 26779 and HIP 113283, bottom). In each plot, the solid colored lines represent the inner habitable zones and the dot-dashed colored lines represent the outer habitable zones. The black vertical dot-dashed lines denote the detector noise corresponding to the HWO EMCCD, the grey vertical dashed lines denote the end-of-life Roman EMCCD detector parameters, and the grey dotted line represents the beginning-of-life Roman EMCCD detector \cite{roman_detect_params}. An energy-resolving detector with 0 detector noise would be off the left-hand side of this plot.}
    \label{fig:etc_res}
\end{figure}

\begin{figure}
    \centering
    \includegraphics[width=0.8\textwidth]{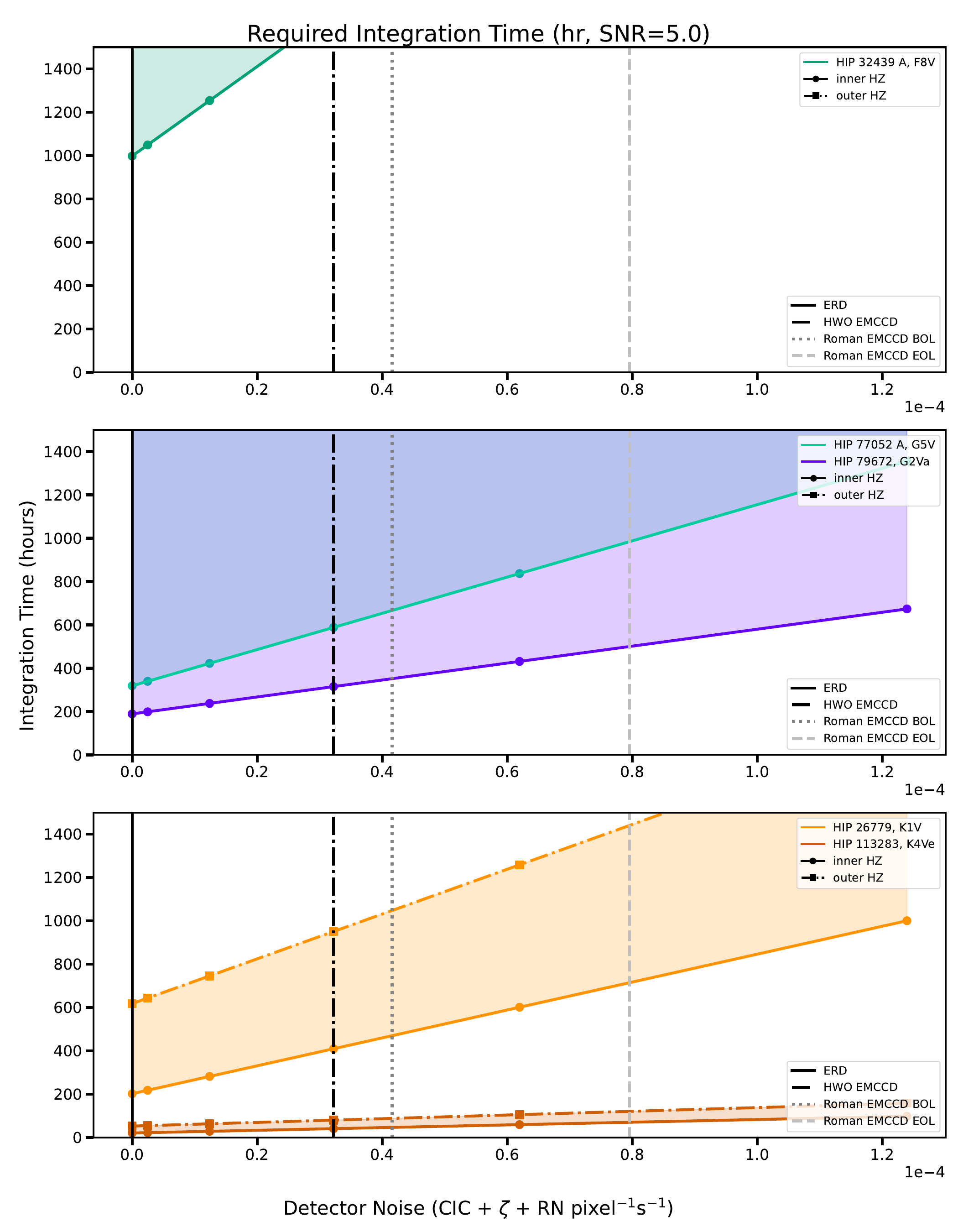}
    \caption{Same as Figure \ref{fig:etc_res}, but zoomed in and put on a linear scale. A solid black line denoting an ERD was also added at 0 detector noise.}
    \label{fig:etc_linear_zoom}
\end{figure}

Here we used EBS to perform a simple case study of how detector noise impacts the time to reach SNR=5 for water detection (R=140) for 5 fiducial HWO target stars as identified by the ExEP Mission Star List for the Habitable Worlds Observatory \cite{2024ExEp_catalog}. A lenslet sampling of 2 was used meaning that the detector noise contribution is coming from spreading the light over 4 pixels per spectral element. The results are found in Figures \ref{fig:etc_res} and \ref{fig:etc_linear_zoom}. The vertical lines denote the detector noises that correspond to the HWO EMCCD (black), the Roman EMCCD beginning-of-life (BOL) performance (dark gray) and the expected Roman end-of-life (EOL) performance (light gray). Figure \ref{fig:etc_linear_zoom} is a zoom-in of Figure \ref{fig:etc_res} with a linear scaling applied to better differentiate the exposure time differences around the detector properties of interest.  

Depending on spectral type, the increase in exposure times required to achieve the same SNR detection moving from an ERD to the HWO EMCCD ranges between factors of 1.5--2.0.

\section{Discussion and Future Work}
\label{sec:discussion}

This work has shown that using an ERD for digging a dark zone could cut the overhead needed to do so by up to a factor of 2 over even a low-noise EMCCD such as the ones budgeted by the LUVOIR and HabEx reports for HWO. Over a mission lifetime, this translates into potentially thousands of hours that could be spent on additional exo-Earth characterization observations (which has been shown in this work and others to also take less time using an ERD) or executing other general astrophysics programs.   

In order to fully quantify the effects of using an ERD in a WFS\&C context beyond what is presented in this work, more detailed detector models and simulations will be required. First, this work assumed that for all EFC iterations, $t_{exp}$ remains constant. In reality, shorter exposures would be required at the beginning of the digging process and get progressively longer as the signal in the dark zone diminishes and so the effect of having adaptive exposure times on contrast is an obvious next step. The parameters used in the EFC (which were fixed for all experiments), such as probe amplitude, and conditioning number, could also be optimized for better overall contrast performance. Additionally, more simulations should be run to more finely sample the range of exposure times and achieve more accurate DZ digging factors as are found in Table \ref{table:time_to_contrast}. For example, since only a finite amount of exposure times were sampled, if a given contrast was just barely not reached for a given $t_{exp}$, then the time to DZ contrast would need to be calculated using the next highest $t_{exp}$ studied (say $t_{exp}$ = 5 min instead of 2 min) whereas, in reality, an exposure time in between those two would yield the most optimal time to a given contrast for that WFS\&C scenario. 

More accurate ERD models should also be used since adapting identically 0 noise for these technologies is neither accurate nor realistic. Adopting values of 0 here has traditionally been an acceptable approximation since the noise for an ERD is significantly lower than that of their semiconducting counterparts, but at these very low photon count rates getting these details correct is important. This study has also been agnostic as to which energy-resolving detector technology is chosen, but adapting more realistic TES or MKID specific detector models should be explored since the output format of ERDs is fundamentally different than semiconductor-based detectors. For example, ERDs are truly photon counting and so the outputs are not integrated images, but photon lists where each photon is tagged with its pixel location, energy, and arrival time that can be operated on entirely in post-processing \cite{2022mkid_pipeline}. Such models will additionally allow the exploration of using photon statistical post-processing techniques in the science ``exposures'' such as those presented in Steiger et. al. (2021) \cite{2021AJsteiger}.


The choice of using an ERD for a future exo-Earth imaging flagship will have wide-reaching impacts on every aspect of the observatory from digging the dark zone, to collecting science data, to post-processing. For this reason, when looking at trade studies involving these detectors, all of these aspects should be taken into consideration to fully quantify their potential benefits.

\clearpage

\clearpage

\clearpage 

\acknowledgments

S.S. is supported by an STScI Postdoctoral Fellowship. E.H.P. was supported in part by the NASA Hubble Fellowship grant \#HST-HF2-51467.001-A awarded by the Space Telescope Science Institute, which is operated by the Association of Universities for Research in Astronomy, Incorporated, under NASA contract NAS5-26555. I.L. acknowledges partial support from a postdoctoral fellowship issued by the Centre National d’Etudes Spatiales (CNES) in France.

The HiCAT testbed has been developed over the past 10 years and benefitted from the work of an extended collaboration of over 50 people. This work was supported in part by the National Aeronautics and Space Administration under Grant 80NSSC19K0120 issued through the Strategic Astrophysics Technology/Technology Demonstration for Exo-planet Missions Program (SAT-TDEM; PI: R. Soummer), and under Grant 80NSSC22K0372 issued through the Astrophysics Research and Analysis Program (APRA; PI: L. Pueyo).

This research made use of HCIPy, an open-source object-oriented framework written in Python for performing end-to-end simulations of high-contrast imaging instruments (Por et al. 2018).
 
\clearpage

\bibliography{main} 

\begin{thebibliography}{10}

\bibitem{PSG}
{Villanueva}, G.~L., {Smith}, M.~D., {Protopapa}, S., {Faggi}, S., and {Mandell}, A.~M., ``{Planetary Spectrum Generator: An accurate online radiative transfer suite for atmospheres, comets, small bodies and exoplanets},'' {\em JQSRT}~{\bf 217},  86--104 (Sept. 2018).

\bibitem{2003Day_mkid}
{Day}, P.~K., {LeDuc}, H.~G., {Mazin}, B.~A., {Vayonakis}, A., and {Zmuidzinas}, J., ``{A broadband superconducting detector suitable for use in large arrays},'' {\em Nature}~{\bf 425},  817--821 (Oct. 2003).

\bibitem{2013ARCONS}
{Mazin}, B.~A., {Meeker}, S.~R., {Strader}, M.~J., {Szypryt}, P., {Marsden}, D., {van Eyken}, J.~C., {Duggan}, G.~E., {Walter}, A.~B., {Ulbricht}, G., {Johnson}, M., {Bumble}, B., {O'Brien}, K., and {Stoughton}, C., ``{ARCONS: A 2024 Pixel Optical through Near-IR Cryogenic Imaging Spectrophotometer},'' {\em PASP}~{\bf 125},  1348 (Nov. 2013).

\bibitem{2018darkness}
{Meeker}, S.~R., {Mazin}, B.~A., {Walter}, A.~B., {Strader}, P., {Fruitwala}, N., {Bockstiegel}, C., {Szypryt}, P., {Ulbricht}, G., {Coiffard}, G., {Bumble}, B., {Cancelo}, G., {Zmuda}, T., {Treptow}, K., {Wilcer}, N., {Collura}, G., {Dodkins}, R., {Lipartito}, I., {Zobrist}, N., {Bottom}, M., {Shelton}, J.~C., {Mawet}, D., {van Eyken}, J.~C., {Vasisht}, G., and {Serabyn}, E., ``{DARKNESS: A Microwave Kinetic Inductance Detector Integral Field Spectrograph for High-contrast Astronomy},'' {\em PASP}~{\bf 130},  065001 (June 2018).

\bibitem{2020MEC}
{Walter}, A.~B., {Fruitwala}, N., {Steiger}, S., {Bailey}, John~I., I., {Zobrist}, N., {Swimmer}, N., {Lipartito}, I., {Smith}, J.~P., {Meeker}, S.~R., {Bockstiegel}, C., {Coiffard}, G., {Dodkins}, R., {Szypryt}, P., {Davis}, K.~K., {Daal}, M., {Bumble}, B., {Collura}, G., {Guyon}, O., {Lozi}, J., {Vievard}, S., {Jovanovic}, N., {Martinache}, F., {Currie}, T., and {Mazin}, B.~A., ``{The MKID Exoplanet Camera for Subaru SCExAO},'' {\em PASP}~{\bf 132},  125005 (Dec. 2020).

\bibitem{2005TES_Irwin}
``{Transition-Edge Sensors},'' in [{\em Cryogenic Particle Detection}{\nolinebreak\hspace{0.1em}]},  {Enss}, C., ed.,  {\bf 99},  63 (2005).

\bibitem{2002SQUID}
{Kiviranta}, M., {Sepp{\"a}}, H., {van der Kuur}, J., and {de Korte}, P., ``{SQUID-based readout schemes for microcalorimeter arrays},'' in [{\em Low Temperature Detectors}{\nolinebreak\hspace{0.1em}]},  {Porter}, F.~S., {McCammon}, D., {Galeazzi}, M., and {Stahle}, C.~K., eds., {\em American Institute of Physics Conference Series} {\bf 605},  295--300, AIP (Feb. 2002).

\bibitem{2024paul_kic}
{Szypryt}, P., {Bennett}, D.~A., {Fogarty Florang}, I., {Fowler}, J.~W., {Giachero}, A., {Hummatov}, R., {Lita}, A.~E., {Mates}, J. A.~B., {Nam}, S.~W., {O'Neil}, G.~C., {Swetz}, D.~S., {Ullom}, J.~N., {Vissers}, M.~R., {Wheeler}, J., and {Gao}, J., ``{Kinetic inductance current sensor for visible to near-infrared wavelength transition-edge sensor readout},'' {\em arXiv e-prints} ,  arXiv:2405.15017 (May 2024).

\bibitem{TES_r}
``Ultra-high efficiency noiseless quantum sensors for hwo and qis..'' \url{https://techport.nasa.gov/view/146757 }.
\newblock Accessed: 2024-06-08.

\bibitem{2021deVisser}
{de Visser}, P.~J., {de Rooij}, S. A.~H., {Murugesan}, V., {Thoen}, D.~J., and {Baselmans}, J. J.~A., ``{Phonon-Trapping-Enhanced Energy Resolution in Superconducting Single-Photon Detectors},'' {\em Physical Review Applied}~{\bf 16},  034051 (Sept. 2021).

\bibitem{2024Howe}
{Howe}, A.~R., {Stark}, C.~C., and {Sadleir}, J.~E., ``{The Scientific Impact of a Noiseless Energy-Resolving Detector for a Future Exoplanet-Imaging Mission},'' {\em arXiv e-prints} ,  arXiv:2405.08883 (May 2024).

\bibitem{2018hicat_software}
{Moriarty}, C., {Brooks}, K., {Soummer}, R., {Perrin}, M., {Comeau}, T., {Brady}, G., {Gontrum}, R., and {Petrone}, P., ``{High-contrast imager for complex aperture telescopes (HiCAT): 6. software control infrastructure and calibration},'' in [{\em Space Telescopes and Instrumentation 2018: Optical, Infrared, and Millimeter Wave}{\nolinebreak\hspace{0.1em}]},  {Lystrup}, M., {MacEwen}, H.~A., {Fazio}, G.~G., {Batalha}, N., {Siegler}, N., and {Tong}, E.~C., eds., {\em Society of Photo-Optical Instrumentation Engineers (SPIE) Conference Series} {\bf 10698},  1069853 (Aug. 2018).

\bibitem{2022Hicat8}
{Soummer}, R., {Por}, E.~H., {Pourcelot}, R., {Redmond}, S., {Laginja}, I., {Will}, S.~D., {Perrin}, M.~D., {Pueyo}, L., {Sahoo}, A., {Petrone}, P., {Brooks}, K.~J., {Fox}, R., {Klein}, A., {Nickson}, B., {Comeau}, T., {Ferrari}, M., {Gontrum}, R., {Hagopian}, J., {Leboulleux}, L., {Leongomez}, D., {Lugten}, J., {Mugnier}, L.~M., {N'Diaye}, M., {Nguyen}, M., {Noss}, J., {Sauvage}, J.-F., {Scott}, N., {Sivaramakrishnan}, A., {Subedi}, H.~B., and {Weinstock}, S., ``{High-contrast imager for complex aperture telescopes (HiCAT): 8. Dark zone demonstration with simultaneous closed-loop low-order wavefront sensing and control},'' in [{\em Space Telescopes and Instrumentation 2022: Optical, Infrared, and Millimeter Wave}{\nolinebreak\hspace{0.1em}]},  {Coyle}, L.~E., {Matsuura}, S., and {Perrin}, M.~D., eds., {\em Society of Photo-Optical Instrumentation Engineers (SPIE) Conference Series} {\bf 12180},  1218026 (Aug. 2022).

\bibitem{SoummerSPIE2024}
{Soummer}, R., {Pourcelot}, R., {Por}, E., {Steiger}, S., {Laginja}, I., {Buralli}, B., {Pueyo}, L., {Nguyen}, M., {Nickson}, B., {Sahoo}, A., and {the extended HiCAT team}, ``High-contrast imager for complex aperture telescopes (hicat): 11. system-level static and dynamic demonstration of the apodized pupil lyot coronagraph with a segmented aperture.,'' {\em Society of Photo-Optical Instrumentation Engineers (SPIE) Conference Series} {\bf 13092} (2024).

\bibitem{2022_APLC_optimization}
{Nickson}, B.~F., {Por}, E.~H., {Nguyen}, M.~M., {Soummer}, R., {Laginja}, I., {Sahoo}, A., {Pueyo}, L., {St. Laurent}, K., {N'Diaye}, M., {Zimmerman}, N.~T., {Noss}, J., and {Perrin}, M., ``{APLC-optimization: an apodized pupil Lyot coronagraph design survey toolkit},'' in [{\em Space Telescopes and Instrumentation 2022: Optical, Infrared, and Millimeter Wave}{\nolinebreak\hspace{0.1em}]},  {Coyle}, L.~E., {Matsuura}, S., and {Perrin}, M.~D., eds., {\em Society of Photo-Optical Instrumentation Engineers (SPIE) Conference Series} {\bf 12180},  121805K (Aug. 2022).

\bibitem{por_2024_catkit}
Por, Emiel, H., Laginja, I., Pourcelot, R., Soummer, R., Sevin, A., Sahoo, A., Nguyen, M., Fowler, J., Egger, L., Pougheon, E., and Demagny, A., ``The control and automation for testbeds kit 2 (catkit2),'' (May 2024).

\bibitem{2019LUVOIR}
{The LUVOIR Team}, ``{The LUVOIR Mission Concept Study Final Report},'' {\em arXiv e-prints} ,  arXiv:1912.06219 (Dec. 2019).

\bibitem{2020HabEx}
{Gaudi}, B.~S., {Seager}, S., {Mennesson}, B., and {et. al.}, ``{The Habitable Exoplanet Observatory (HabEx) Mission Concept Study Final Report},'' {\em arXiv e-prints} ,  arXiv:2001.06683 (Jan. 2020).

\bibitem{roman_detect_params}
``Nancy grace roman space telescope simulations, spacecraft and instrument parameters..'' \url{https://roman.ipac.caltech.edu/sims/Param_db.html }.
\newblock Accessed: 2024-05-31.

\bibitem{2007EFC}
{Give'on}, A., {Kern}, B., {Shaklan}, S., {Moody}, D.~C., and {Pueyo}, L., ``{Broadband wavefront correction algorithm for high-contrast imaging systems},'' in [{\em Astronomical Adaptive Optics Systems and Applications III}{\nolinebreak\hspace{0.1em}]},  {Tyson}, R.~K. and {Lloyd-Hart}, M., eds., {\em Society of Photo-Optical Instrumentation Engineers (SPIE) Conference Series} {\bf 6691},  66910A (Sept. 2007).

\bibitem{2017EXOSIMS}
{Savransky}, D., {Delacroix}, C., and {Garrett}, D., ``{EXOSIMS: Exoplanet Open-Source Imaging Mission Simulator}.'' Astrophysics Source Code Library, record ascl:1706.010 (June 2017).

\bibitem{2024ExEp_catalog}
{Mamajek}, E. and {Stapelfeldt}, K., ``{Provisional NASA ExEP Mission Target Star List for the Habitable Worlds Observatory},'' in [{\em AAS/Division for Extreme Solar Systems Abstracts}{\nolinebreak\hspace{0.1em}]},  {\em AAS/Division for Extreme Solar Systems Abstracts} {\bf 56},  628.17 (Apr. 2024).

\bibitem{2024HPIC}
{Tuchow}, N.~W., {Stark}, C.~C., and {Mamajek}, E., ``{HPIC: The Habitable Worlds Observatory Preliminary Input Catalog},'' {\em AJ}~{\bf 167},  139 (Mar. 2024).

\bibitem{2020Nematib}
{Nemati}, B., ``{Photon counting and precision photometry for the Roman Space Telescope Coronagraph},'' in [{\em Space Telescopes and Instrumentation 2020: Optical, Infrared, and Millimeter Wave}{\nolinebreak\hspace{0.1em}]},  {Lystrup}, M. and {Perrin}, M.~D., eds., {\em Society of Photo-Optical Instrumentation Engineers (SPIE) Conference Series} {\bf 11443},  114435F (Dec. 2020).

\bibitem{por2018hcipy}
Por, E.~H., Haffert, S.~Y., Radhakrishnan, V.~M., Doelman, D.~S., Van~Kooten, M., and Bos, S.~P., ``{High Contrast Imaging for Python (HCIPy): an open-source adaptive optics and coronagraph simulator},'' in [{\em Adaptive Optics Systems VI}{\nolinebreak\hspace{0.1em}]},  {\em Proc. {{SPIE}}} {\bf 10703} (2018).

\bibitem{2020nemati_coronagraphs}
{Nemati}, B., {Stahl}, H.~P., {Stahl}, M.~T., {Ruane}, G.~J., and {Sheldon}, L.~J., ``{Method for deriving optical telescope performance specifications for Earth-detecting coronagraphs},'' {\em Journal of Astronomical Telescopes, Instruments, and Systems}~{\bf 6},  039002 (July 2020).

\bibitem{2022mkid_pipeline}
{Steiger}, S., {Bailey}, J.~I., {Zobrist}, N., {Swimmer}, N., {Dodkins}, R., {Davis}, K.~K., and {Mazin}, B.~A., ``{The MKID Pipeline: A Data Reduction and Analysis Pipeline for UVOIR MKID Data},'' {\em AJ}~{\bf 163},  193 (May 2022).

\bibitem{2021AJsteiger}
{Steiger}, S., {Currie}, T., {Brandt}, T.~D., {Guyon}, O., {Kuzuhara}, M., {Chilcote}, J., {Groff}, T.~D., {Lozi}, J., {Walter}, A.~B., {Fruitwala}, N., {Bailey}, John~I., I., {Zobrist}, N., {Swimmer}, N., {Lipartito}, I., {Smith}, J.~P., {Bockstiegel}, C., {Meeker}, S.~R., {Coiffard}, G., {Dodkins}, R., {Szypryt}, P., {Davis}, K.~K., {Daal}, M., {Bumble}, B., {Vievard}, S., {Sahoo}, A., {Deo}, V., {Jovanovic}, N., {Martinache}, F., {Doppmann}, G., {Tamura}, M., {Kasdin}, N.~J., and {Mazin}, B.~A., ``{SCExAO/MEC and CHARIS Discovery of a Low-mass, 6 au Separation Companion to HIP 109427 Using Stochastic Speckle Discrimination and High-contrast Spectroscopy},'' {\em AJ}~{\bf 162},  44 (Aug. 2021).

\end{thebibliography}
\bibliographystyle{spiebib} 

\end{document}